\documentclass[amsmath,amssymb]{revtex4}

\usepackage{graphicx}
\usepackage{bm}

\def \tx{\tilde{x}}
\def \tc{\tilde{c}}
\def \tm{\tilde{m}}
\def \tM{\tilde{M}}
\def \tU{\tilde{U}}

\def \tz{\tilde{z}}
\def \tZ{\tilde{Z}}
\def \tr{\tilde{r}}
\def \tq{\tilde{q}}
\def \txi{\tilde{\xi}}
\def \tY{\tilde{Y}}
\def \tSigma{\tilde{\Sigma}}
\def \Prob{{\mathrm {Prob}}}
\def\Vec#1{\boldsymbol {{#1}}}

\begin{document}
\sloppy

\title{Analysis of Bidirectional Associative Memory using SCSNA and Statistical Neurodynamics}
\author{Hayaru Shouno}
\affiliation{Dept. of Computer Science and Systems Engineering, Faculty of Engineering, Yamaguchi University}
\email{shouno@ai.csse.yamaguchi-u.ac.jp}

\author{Shoji Kido}
\affiliation{Dept. of Computer Science and Systems Engineering, Faculty of Engineering, Yamaguchi University}

\author{Masato Okada}
\affiliation{Brain Research Institute, RIKEN}

\date{\today}


\begin{abstract}
Bidirectional associative memory (BAM) is a kind of 
an artificial neural network used to memorize and retrieve heterogeneous pattern pairs.
%
Many efforts have been made to improve BAM from the 
the viewpoint of computer application, 
and
few theoretical studies have been done.
We investigated the theoretical characteristics of BAM using 
a framework of statistical-mechanical analysis.
To investigate the equilibrium state of BAM, 
we applied self-consistent signal to noise analysis (SCSNA) and 
obtained a macroscopic parameter equations and relative capacity.
%
%
Moreover, to investigate not only the equilibrium state but also 
the retrieval process of reaching the equilibrium state,
we applied statistical neurodynamics to the update rule of BAM
and obtained evolution equations for the macroscopic parameters.
These evolution equations are consistent with the results of 
SCSNA in the equilibrium state.
\end{abstract}

\maketitle
\section{Introduction}
Bi-directional associative memory (BAM) \cite{Kosko88} is a kind of
an associative memory model which is an artificial neural network.
The principle function of associative memory is 
to memorize multiple patterns and to retrieve the correct one
when a pattern key is given.

Autocorrelation associative memory (AAM), 
sometimes called the Hopfield model \cite{Hopfield86},
is also a kind of associative memory. 
AAM tries to retrieve a stored pattern when 
a degraded pattern is given as an association key;
this type of retrieving is called homogeneous association.
In contrast, BAM stores multiple pattern pairs 
and 
tries to retrieve a complete stored pattern pair 
when 
a degraded piece of the pair is given as an association key.
Thus, BAM is called a heterogeneous pattern association model.

In the field of neural networks, 
many efforts have been made to improve BAM from the viewpoint of
computer application \cite{Hassoun89} \cite{Simpson90} \cite{Wang90}
\cite{Zhuang93} \cite{Oh94} \cite{Wang95} \cite{Wang96} \cite{Hongchi98},
and few theoretical analyses have been reported \cite{Haines88} \cite{Yanai91} \cite{Tanaka00}.
The theoretical analysis of BAM has evolved with a focus on storage capacity, 
which means how many patterns can be stored in a network 
consisted of $N$ neural units.
Yanai {\it et al.} suggested that 
BAM can be regarded as a variation of AAM 
in which connections are systematically reduced \cite{Yanai91}.
They also showed that the relative storage capacity,
in which a finite amount of retrieval error is allowed,
of BAM to be around $0.22 N$.
Haines \& Hecht-Nielsen analyzed BAM along the same way, 
and reported its absolute capacity, 
in which no retrieval error is allowed, 
to be $O(N/\log N)$ \cite{Haines88}.
Tanaka {\it et al.} analyzed BAM using a replica method
(see \cite{FisherHertz91}), 
which is a statistical-mechanical analysis method,
and showed its relative capacity to be 
$0.1998 N$ \cite{Tanaka00}.
These analyses mainly focused on the equilibrium state of BAM, 
and 
the transient process of retrieving, 
which means how to reach the equilibrium state,
was not so conducted.
However, analysis of the retrieval process is as important as
that of the equilibrium state.

In this paper, we have analyzed the equilibrium state of BAM using
the self-consistent signal-to-noise analysis (SCSNA)\cite{Shiino92}.
We found that the relative capacity was $0.1998N$, 
which agrees with the result of Tanaka {\it et al}.
We also investigated the retrieval process of BAM;
we derived macroscopic dynamical equations using 
the statistical neurodynamics, 
which was theoretically derived in the same manner as the SCSNA
\cite{Amari88a} \cite{Okada95},
and compared the results of between the statistical neurodynamics with
those of computer simulation.
Applying the statistical neurodynamics to BAM, 
we obtained the evolution equations for the macroscopic parameters.
In the limit of these evolution equations, 
that is, macroscopic parameters of BAM reached the equilibrium state, 
we found these values were consistent with the results of SCSNA.
We also compared the results of applying the statistical
neurodynamics with those of the computer simulation
and obtained quantitative support for our analysis.

%

We describe the BAM formulation in Section
\ref{sec:formulation}, 
and we apply the SCSNA and show the results of 
equilibrium state analysis in Section \ref{sec:SCSNA}.
In Section \ref{sec:dynamics}, 
we derived the evolution equations of macroscopic parameters using
the statistical neurodynamics
and compared the results with those of computer
simulation in Section \ref{sec:compare}.

\section{Formulation}
\label{sec:formulation}
\begin{figure}
 \begin{center}
  \resizebox{7.8cm}{!}{\includegraphics{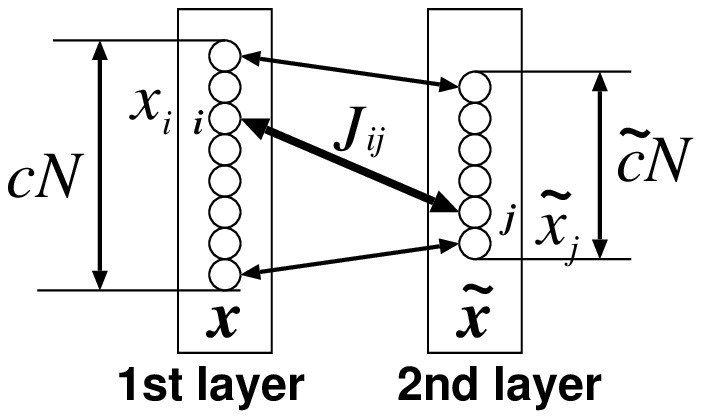}}
  \caption{Network structure of BAM}
  \label{fig:bam}
 \end{center}
\end{figure}

As shown in Fig. \ref{fig:bam},
BAM is a two-layered neural network model \cite{Kosko88}. 
The first layer consists of $c N$  neural units ($c\sim O(1)$), 
and the state of the layer is denoted as $\Vec{x}$
with the components denoted as $x_i\:\:(1\leq i \leq cN)$.
The state of the second layer, which has $\tc N$ units, is denoted as
$\Vec{\tx}$,
and the $j$th unit state is described as $\tx_j \:\:(1\leq j \leq\tc N)$.
Each layer is connected by interlayer connection $\Vec{J}$ 
with the components described as 
$J_{ij} \:\:(1\leq i \leq cN, \: 1 \leq j \leq \tc N)$.
$J_{ij}$ represents the connection weight between 
the first layer unit $x_i$ and 
the second layer unit ${\tx}_j$.

We prepare $p$ binary pattern pairs denoted as 
$\{\Vec{\xi}^{\mu}, \Vec{\txi}^{\mu}\}$ $(\mu = 1, \cdots, p)$,
where the superscript $\mu$ denotes the pattern pair index.
Pattern vector $\Vec{\xi}^{\mu}$ corresponds to the first layer, and
$\Vec{\txi}^\mu$ corresponds to the second layer.
Thus $\Vec{\xi}^{\mu}$ and $\Vec{\txi}^{\mu}$ have $c N$ and $\tc N$
components, respectively, and each component,
which is described as 
$\xi^{\mu}_i$ and ${\txi}^{\mu}_j$
$(1 \leq i \leq c N, \:\: 1 \leq j \leq \tc N)$, respectively,
is generated 
from uniform i.i.d.:
\begin{align}
 \Prob[\xi_i^{\mu}=\pm 1] &= \frac{1}{2}, \label{eq:pat1}\\
 \Prob[\txi_j^{\mu} \pm 1] &=  \frac{1}{2} \label{eq:pat2}.
\end{align}
Assuming the number of stored pattern pairs to be $p\sim O(N)$, 
we define a quantity $\alpha \left( = \frac{p}{N} \right)$,
and use it for the loading rate.
%
%

To determine the interlayer connection weight $J_{ij}$, which connects
$x_i$ and $\tx_j$, we use a correlation-based learning rule:
\begin{equation}
 J_{ij} = \frac{1}{N} \sum_{\mu=1}^{\alpha N} \xi^{\mu}_{i} \txi^{\mu}_{j}.
\label{eq3}
\end{equation}
%
%
All the pattern pair correlations between $\Vec{\xi}^{\mu}$ and
$\Vec{\txi}^{\mu}$ are embedded in connection weight $\Vec{J}$.
In this notation, 
the connections are not symmetrical, that is $J_{ij} \neq J_{ji}$.

In the retrieving, 
we use a synchronous update rule for each layer;
that is, all units in each layer are updated synchronously,
and these layers are updated alternately.
The rules for updating the $i$th unit in the first layer
and the $j$th unit in the second layer are
\begin{align}
  x_{i}^{2t} &= F( \sum_{j=1}^{\tc N} J_{ij} \tx_{j}^{2t-1}  ), 
  \label{eq:dynamics1} \\
 \tx_{j}^{2t+1} &= F( \sum_{i=1}^{cN} J_{ij} x_{i}^{2t} ) 
  \label{eq:dynamics2},
\end{align}
where $t$ means the one step time and $F(\cdot)$ means the output function.
%
In these formulations, the retrieval process is carried out as follows.
In a initial state, $t=0$, 
Association key $\Vec{x}^{0} (= \{ x_i^{0} \})$ is given to the first layer.
Then, 
all of the second layer units, $\tx_{j}^{1} \: (1\leq j \leq \tc N)$,
are updated using eq. (\ref{eq:dynamics2}), and 
the state of the second layer is described as $\Vec{\tx}^{1}$.
Next, $t=1$, all the units in the first layer, $x_{i}^{2} \: (1\leq i \leq c N)$, 
are updated using eq.(\ref{eq:dynamics1}),
and the state is described as $\Vec{x}^{2}$.
After that the second layer is updated by eq.(\ref{eq:dynamics2}).
For each $t = 2, 3, \cdots$, the alternate updating of each layer
are carried out in the same way, 
and each layer state is denoted as $\Vec{x}^{2t}$ and $\Vec{\tx}^{2t+1}$.

To apply S/N analysis, we introduce overlaps, which means similarities between patterns. 
The overlaps between first layer state $\Vec{x}^{2t}$ 
and 
the $\mu$th pattern, $\Vec{\xi}^{\mu}$, 
and 
between second-layer state $\Vec{\tx}^{2t-1}$ and 
$\Vec{\txi}^{\mu}$
are described as follows, respectively:
\begin{align}
 m^{2t}_{\mu}   &= \frac{1}{cN} \sum_{i=1}^{cN} x^{2t}_{i} \xi^{\mu}_{i}, \\
 \tm^{2t+1}_{\mu} &= \frac{1}{\tc N} \sum_{j=1}^{\tc N} \tx^{2t+1}_{j} \txi^{\mu}_{j}.
\end{align}
Following the S/N analysis,
we decomposed the inner term of $F(\cdot)$ in eqs.(\ref{eq:dynamics1}) and (\ref{eq:dynamics2}) 
into signal and noise components.
Assuming the first pattern pair, $\{ \Vec{\xi}^1, \Vec{\txi}^1 \}$,
is retrieved, 
the terms including overlaps $m_1$ and $\tm_1$, 
are signal components, {\it i.e.}  $m_1$, $\tm_1$ $\sim O(1)$. 
Using these overlaps, 
eqs.(\ref{eq:dynamics1}) and (\ref{eq:dynamics2}) can be described as
\begin{align}
 x^{2t}_{i} &= F(  \tc \tm^{2t-1}_1 \xi^{1}_{i} + z^{2t-1}_i ),
 \label{eq:ov_update1}\\
 \tx^{2t+1}_{j} &= F( c m^{2t}_1 \txi^{1}_j + \tz^{2t}_j ), 
 \label{eq:ov_update2}
\end{align}
where $z^{2t-1}_i$, and $\tz^{2t}_j$ are called as crosstalk noises, 
which 
prevents the target pair $\{ \Vec{\xi}^1, \Vec{\txi}^1\}$ to be retrieved.
These crosstalk noises are denoted
\begin{align}
 z^{2t-1}_i &= \frac{1}{N} \sum_{\mu=2}^{\alpha N} \sum_{j=1}^{\tc N}
 \xi^{\mu}_i \txi^{\mu}_j \tx^{2t-1}_j,  
 \label{eq:noise1}\\
 \tz^{2t}_j &= \frac{1}{N} \sum_{\mu=2}^{\alpha N} \sum_{i=1}^{cN}
 \txi^{\mu}_j \xi^{\mu}_i x^{2t}_i.
 \label{eq:noise2}
\end{align}

\section{Equilibrium state analysis by SCSNA}
\label{sec:SCSNA}
To derive equilibrium state macroscopic parameters, 
we use SCSNA\cite{Shiino92}, which is an extension of a naive 
signal-to-noise (S/N) analysis.
Since the SCSNA treats the equilibrium states of an associative memory model,
we omit the index $t$ in the update rules.
%
Hence we can rewrite eqs. (\ref{eq:ov_update1})  and (\ref{eq:ov_update2}) as
\begin{align}
 x_{i} &= F(  \tc \tm_1 \xi^{1}_{i} + z_i ),
 \label{eq:eq_update1}
 \\
 \tx_{j} &= F( c m_1 \txi^{1}_j + \tz_j ), 
 \label{eq:eq_update2}
\end{align}
respectively.
%

\begin{figure}[t]
 \begin{center}
  \resizebox{0.8\textwidth}{!}{\includegraphics{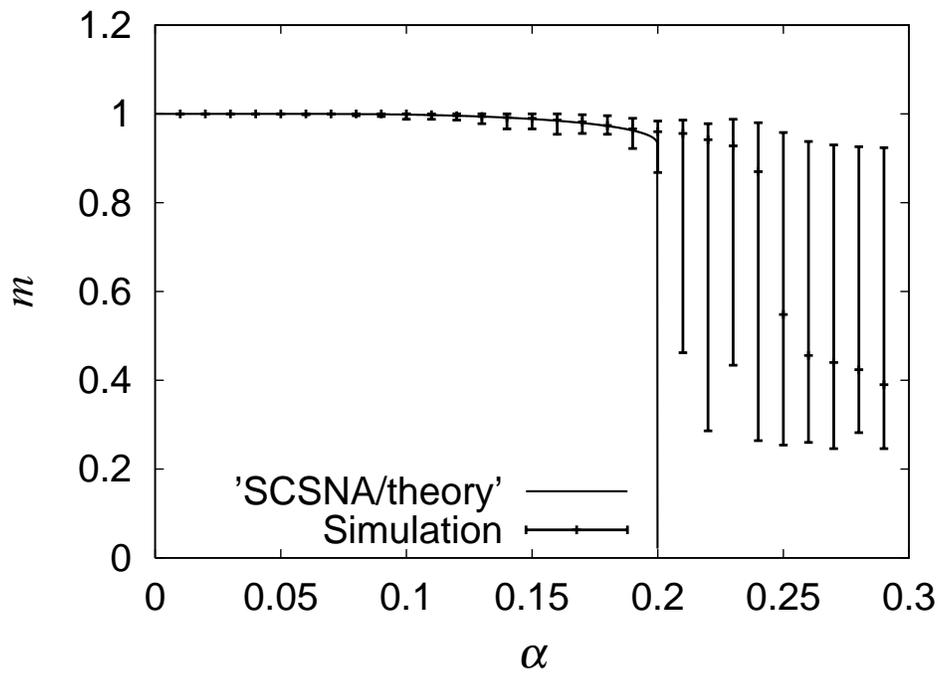}}
 \end{center}
 \caption{Comparing SCSNA results with those of the
 simulation. The horizontal axis means the loading rate $\alpha$, and the
 vertical axis means the overlap. The results of computer simulation are
 shown as error-bars, which indicates median with minimum and maximum values.}
 \label{fig:fig1}
\end{figure}

%
%
In the SCSNA, the crosstalk noise term is decomposed into 
a systematic bias term and 
a Gaussian noise term with $0$ mean\cite{Shiino92}.
The detailed formulas of SCSNA are described in appendix.
We derive
self-consistent equations called order parameter equations. 
The following are the order parameter equations of BAM.
\small
\begin{align}
 Y &= F( \tc \tm \xi  + 
  \frac{\alpha \tc \tU}{1-c\tc U\tU} Y +
  \sqrt{\alpha r} z ), \label{eq:param_from}\\
 \tY &= F( cm \txi + 
  \frac{\alpha c U}{1-c\tc U\tU} \tY +
  \sqrt{\alpha \tr} z ), \\
 m &= \int Dz \: \langle \xi Y \rangle_{\xi}, \\
 \tm &= \int Dz \langle \txi \tY \rangle_{\txi}, \\
 q &= \int Dz \langle Y^2 \rangle_{\xi}, \\
 \tq &= \int Dz \langle \tY^2 \rangle_{\txi}, \\
 U &= \frac{1}{\sqrt{\alpha r}}
  \int Dz z \langle Y \rangle_{\xi}, \\
 \tU &= \frac{1}{\sqrt{\alpha \tr}}
  \int Dz z \langle \tY \rangle_{\txi}, \\
 r &= \frac{\tc}{(1-c\tc U\tU)^2} (\tq + c\tc \tU^2 q),\\
 \tr &= \frac{c}{(1-c\tc U\tU)^2} (q + c\tc U^2 \tq).
  \label{eq:param_to}
\end{align}
\normalsize
These equations are described in the manner of Shiino and Fukai \cite{Shiino92}. 
$Y$ and $\tY$ represents 
the effective outputs for $x_i$ and $\tx_j$, respectively.
The stochastic variables $\xi$ and $\txi$, 
obeying eq.(\ref{eq:pat1}) and (\ref{eq:pat2}), 
corresponds to a retrieving pattern components $\xi^1_i$ and $\txi^1_j$,
and order parameters $m$ and $\tm$ corresponds to overlaps $m_1$ and
$\tm_1$.
Note that the operators $\langle\cdot\rangle_{\xi}$ and
$\langle\cdot\rangle_{\txi}$ mean the 
expectations for stochastic variables $\xi$ or $\txi$, respectively.
%
%
%
Each arguments of the function $F(\cdot)$ consists of three parts.
The first terms, $\tc\tm\xi$ and $cm\txi$, come from the signal components, 
the second terms, $\frac{\alpha \tc \tU}{1-c\tc U\tU} Y$ and 
$\frac{\alpha c U}{1-c\tc U\tU} \tY$, mean the systematic bias of 
the crosstalk noises ($z_i$,  $\tz_j$) in eqs. (\ref{eq:noise1}) and
(\ref{eq:noise2}),    
and 
each third term is assigned to be a Gaussian distribution with
$0$ mean and  $\alpha r$ or $\alpha \tr$ variance.
%
%
%
We solved the order parameter equations from (\ref{eq:param_from}) to
(\ref{eq:param_to}) numerically and compared the results with those of
simulations. 
Fig. \ref{fig:fig1}. shows the equilibrium overlap $m$ against
the capacity parameter $\alpha$.
An overlap of $1$ means that 
the BAM retrieves a stored pattern pair successfully. 
We obtained a relative capacity, $\alpha_c$ of $0.1998$ 
in which the nontrivial solution $m \neq 0$ and $\tm \neq 0$ is disappeared.
This agrees with the results of Tanaka {\it et al.} ($\alpha_c =
0.1998$),
obtained with the replica method \cite{Tanaka00}.
In fig.\ref{fig:fig1}, we show the simulation results as error-bars, 
which mean  medians and quartile deviations for ten trials. 
The SCSNA results quantitatively explained the simulation results very well.

\section{Retrieval process of BAM}
\label{sec:dynamics}
As we have seen, the SCSNA described the equilibrium state of BAM
quantitatively.  
In this section, we consider a retrieval process of BAM, which means 
the transient process reaching the equilibrium state.
The statistical neurodynamics, 
which is a theory for the retrieval for associative memory model,
is based on S/N analysis.
Amari and Maginu proposed a statistical neurodynamical theory on 
the S/N analysis\cite{Amari88a}, 
%
%
It was known that 
the storage capacity obtained by Amari \& Maginu theory does not
coincide with the results of the replica theory\cite{Amit85b},
and 
the size of the basin of attraction derived from Amari \& Maginu theory
is larger than the results of the computer simulation.
Okada extended the Amari \& Maginu theory to improve to resolve these
difficulties\cite{Okada95},
and obtained a macroscopic equation which has hierarchical structure.
In the macroscopic equation,
the first-order approximation corresponds to the Amari \& Maginu theory, 
and
the higher order approximation coincide with the replica theory.
%
%
%

For applying the statistical neurodynamics to BAM,  
we evaluate the crosstalk noises (eqs.(\ref{eq:noise1}) and
(\ref{eq:noise2})) in eqs. (\ref{eq:ov_update1}) and
(\ref{eq:ov_update2}).
Assuming the first pattern pair, $\{\Vec{\xi}^1,  \Vec{\txi}^1 \}$, 
is retrieved,  
we can regard the overlaps of other pattern pairs,
$\{m^{2t}_{\mu}, \tm^{2t+1}_{\mu}\}$ where $\mu \geq 2$,
as small.
Thus, we expand the state $x_i^{2t}$ and $\tx_j^{2t+1}$:
\begin{align}
 x_i^{2t} &= 
 x_i^{2t,(\mu)}
 + \tc \tm^{2t-1}_\mu \xi_i^{\mu}
 F'(\tc \sum_{\nu \neq \mu}^{\alpha N} \tm^{2t-1}_{\mu} \xi_i^{\nu} ),
 \label{eq:expand1}
 \\
 \tx_j^{2t+1} &= 
 \tx_j^{2t+1, (\mu)}
 + c m^{2t}_\mu \txi_j^{\mu}
 F'(c \sum_{\nu \neq \mu}^{\alpha N} m^{2t}_{\nu} \txi_j^{\nu} ),
 \label{eq:expand2}
\end{align}
for $\mu \geq 2$, where
$\tx_j^{2t+1, (\mu)} = 
F(c \displaystyle \sum_{\nu \neq \mu}^{\alpha N} m^{2t}_{\nu} \txi_j^{\nu} )$ 
and 
$x_i^{2t,(\mu)} = 
F(\tc \displaystyle \sum_{\nu \neq \mu}^{\alpha N} \tm^{2t-1}_{\mu} \xi_i^{\nu} )$.
Substituting eqs. (\ref{eq:expand1}) and (\ref{eq:expand2}) into 
eqs.(\ref{eq:noise1}) and (\ref{eq:noise2}), we obtain,
\begin{align}
 z^{2t+1}_i &= \alpha \tc \tU_{2t+1} x_i^{2t,(\mu)} + Z_i^{2t+1},
\\
 \tz^{2t}_j &= \alpha c U_{2t} \tx_j^{2t-1,(\mu)} + \tZ_j^{2t},
\end{align}
where
\begin{align}
 Z_i^{2t+1} &= \frac{1}{N} 
 \sum_{\mu=2}^{\alpha N} \sum_{j=1}^{\tc N}
 \xi^{\mu}_i \txi^{\mu}_j \tx^{2t+1,(\mu)}_j +
 \frac{\tc \tU_{2t+1}}{N} \sum_{\mu=2}^{\alpha N} \sum_{k\neq i}^{c N}
 \xi^{\mu}_i \xi^{\mu}_k x^{2t,(\mu)}_k +
 \tc c \tU_{2t+1} U_{2t} Z^{2t-1}_i,
 \label{eq:noise_expand1}
 \\
 \tZ_j^{2t} &= \frac{1}{N} 
 \sum_{\mu=2}^{\alpha N} \sum_{i=1}^{c N}
 \txi^{\mu}_j \xi^{\mu}_i x^{2t,(\mu)}_i +
 \frac{c U_{2t}}{N} \sum_{\mu=2}^{\alpha N} \sum_{l\neq j}^{\tc N}
 \txi^{\mu}_j \txi^{\mu}_l \tx^{2t-1,(\mu)}_l +
 c \tc U_{2t} \tU_{2t-1} \tZ^{2t-2}_j,
 \label{eq:noise_expand2}
\end{align}
where
\begin{align}
 \tU_{2t+1} &= \frac{1}{\tc N}  
 \sum_{j=1}^{\tc N} 
 F'(c \sum_{\nu \neq \mu}^{\alpha N} m^{2t}_{\nu} \txi_j^{\nu} ) ,\\
 U_{2t} &= \frac{1}{c N}  
 \sum_{i=1}^{c N}  
 F'(\tc \sum_{\nu \neq \mu}^{\alpha N} \tm^{2t-1}_{\mu} \xi_i^{\nu} ).
\end{align}
Since $x_i^{2t,(\mu)}$ and $\tx_j^{2t-1,(\mu)}$ are almost independent
with $\xi_i^{\mu}$ and $\txi_j^{\mu}$, respectively,
each $Z_i^{2t+1}$  and $\tZ_j^{2t}$ can be regarded as 
independent identical Gaussian distributions, 
that is $Z_i^{2t+1} \sim N(0, \alpha r_{2t+1})$ and 
$\tZ_j^{2t} \sim N(0, \alpha \tr_{2t})$.
Each noise variance,  
$E[(Z_i^{2t+1})^2] = \alpha r_{2t+1}$ and 
$E[(\tZ_j^{2t})^2] = \alpha \tr_{2t}$, 
can be described as
\begin{align}
 \alpha r_{2t+1} &=
 \alpha \tc \tq_{2t+1} +
 \alpha c \tc^2 \tU_{2t+1}^2 q_{2t} +
 \alpha (\tc c \tU_{2t+1} U_{2t})^2 r_{2t-1}  \notag\\
 & \:\:\:\:
 + 2 c\tc \tU_{2t+1} U_{2t}
 E\left[
 Z^{2t-1}_i  
 \frac{1}{N}
 \sum_{\mu=2}^{\alpha N} \sum_{j=1}^{\tc N}
 \xi^{\mu}_i \txi^{\mu}_j \tx^{2t+1,(\mu)}_j
 \right]
 \label{eq:r_expand1}
 ,
 \\
 \alpha \tr_{2t} &=
 \alpha c q_{2t} +
 \alpha \tc c^2 U_{2t}^2 \tq_{2t-1} +
 \alpha (c \tc U_{2t} \tU_{2t-1})^2 \tr_{2t-2} \notag\\
 & \:\:\:\:
  + 2 c\tc U_{2t} \tU_{2t-1}
  E\left[
    \tZ^{2t-2}_j
    \frac{1}{N}
    \sum_{\mu=2}^{\alpha N} \sum_{i=1}^{c N}
    \txi^{\mu}_j \xi^{\mu}_i x^{2t,(\mu)}_i
  \right],
 \label{eq:r_expand2}
\end{align}
where
\begin{align}
 \tq_{2t+1} &= \frac{1}{\tc N}
 \sum_{j=1}^{\tc N}  \left( \tx_j^{2t+1,(\mu)} \right)^2,
 \\
 q_{2t} &= \frac{1}{c N}
 \sum_{i=1}^{c N} \left( x_i^{2t,(\mu)} \right)^2.
\end{align}
The last terms in eqs.(\ref{eq:r_expand1}) and (\ref{eq:r_expand2}) are
determined  by correlations between 
the current state $\tx_j^{2t+1}$ and the previous state noise variable
$Z_i^{2t-1}$,
and between $x_i^{2t}$ and $\tZ_j^{2t-2}$, respectively.
Assuming that the $(n+1)$ previous state noise variables
$Z_i^{2(t-n)-1}$ and $\tZ_j^{2(t-n)-2}$ have no correlation with the 
current state $\tx_j^{2t+1}$ and $x_i^{2t}$, respectively,
we can expand $r_{2t+1}$ and $\tr_{2t}$ as recurrence formulas:
\begin{align}
 r_{2t+1} &= \tc \tq_{2t+1} + c (\tc\tU_{2t+1})^2 q_{2t}  
 + (c\tc \tU_{2t+1} U_{2t})^2 r_{2t-1} \notag\\
  & 
   +
   2\tc \sum_{\eta=1}^{n}
  (c\tc)^{\eta} 
  \tq_{2t+1,2(t-\eta)+1}
  \!\!\!\!
  \prod_{\tau=t-\eta+1}^{t}
  \!\!\!\!
  \tU_{2\tau+1} U_{2\tau} 
 \notag \\
  & 
  +
  2c (\tc\tU_{2t+1})^2 \sum_{\eta=1}^{n-1}
  (c\tc)^{\eta} 
  q_{2t,2(t-\eta)}
  \!\!\!\!\!
  \prod_{\tau=t-\eta+1}^{t}
  \!\!\!\!\!
  U_{2\tau} \tU_{2\tau-1} 
 \label{eq:r_expand1d},
\end{align}
\begin{align}
 \tr_{2t} &= c q_{2t} + \tc (cU_{2t}^2)^2 \tq_{2t-1} 
 + (\tc c U_{2t} \tU_{2t-1})^2 \tr_{2t-2},
 \notag\\
 & 
  + 
  2 c \sum_{\eta=1}^{n}
  (\tc c)^{\eta} 
  q_{2t,2(t-\eta)}
  \!\!\!\!
  \prod_{\tau=t-\eta+1}^{t}
  \!\!\!\!
  U_{2\tau} \tU_{2\tau-1} 
 \notag\\
 &
  +
  2 \tc (cU_{2t})^2
  \sum_{\eta=1}^{n-1}
  (\tc c)^{\eta} 
  \tq_{2t-1,2(t-\eta)-1}
  \!\!\!\!\!\!\!\!
  \prod_{\tau=t-\eta+1}^{t}
  \!\!\!\!\!\!
  \tU_{2\tau-1} U_{2\tau-2},
 \label{eq:r_expand2d}
\end{align}
where $\tq_{2t+1,2(t-n)+1}$ means a cross-correlation between
the current state $\tx_j^{2t+1}$ and the n-step previous state
$\tx_j^{2(t-n)+1}$, 
and $q_{2t, 2(t-n)}$ means a cross-correlation between $x_i^{2t}$ and 
$x_i^{2(t-n)}$.
These variables can be also described with the macroscopic parameters across 
the n-step previous state.
The complete formula is described in the appendix.
%

We obtain the evolution equations
for macroscopic parameters as follows:
\begin{align}
 Y^{2t} &= F( \tc \tm_{2t-1} \xi + \sqrt{\alpha r_{2t-1}} z ), 
 \label{eq:dyn_start}
 \\
 \tY^{2t+1} &= F( c m_{2t} \txi  + \sqrt{\alpha \tr_{2t}} z ), \\
 m^{2t} &= \int Dz \langle \xi Y^{2t} \rangle_{\xi}, \\
 \tm^{2t+1} &= \int Dz \langle \txi \tY^{2t+1} \rangle_{\txi},\\
 q_{2t} &= \int Dz \langle (Y^{2t})^2 \rangle_{\xi}, \\
 \tq_{2t+1} &= \int Dz \langle (\tY^{2t+1})^2 \rangle_{\txi},
\end{align}
\begin{align}
 U_{2t} &= \frac{1}{\sqrt{\alpha r_{2t-1}}}
  \int Dz z \langle Y^{2t} \rangle_{\xi}, \\
 \tU_{2t+1} &= \frac{1}{\sqrt{\alpha \tr_{2t}}}
  \int Dz z \langle \tY^{2t+1} \rangle_{\txi} 
 \label{eq:dyn_end}
\end{align}
In these order parameter equations, 
$Y^{2t}$ and $\tY^{2t+1}$ correspond to the $x_i^{2t}$
and $\tx_j^{2t+1}$, respectively.
The overlaps for the first pattern pair, $m^{2t}_1$ and $\tm^{2t+1}_1$, 
which mean retrieval degree, correspond to the $m^{2t}$ and $\tm^{2t+1}$,
respectively.

Yanai {\it et al.} \cite{Yanai91} applied the one-step analysis,
which corresponds to Amari \& Maginu theory.
In their analysis, 
the macroscopic order parameter equations for
$Y_i^{2t}$, $\tY_j^{2t+1}$, $m^{2t}_1$, $\tm^{2t+1}_1$, 
$q_{2t}$, $\tq_{2t+1}$, $U_{2t}$,$\tU_{2t+1}$, 
which are described in eqs. from (\ref{eq:dyn_start})
 to (\ref{eq:dyn_end}),
are identical to those of our analysis.
The differences are in evaluating of noise variances,
that is, $r_{2t+1}$ and $\tr_{2t}$.
They ignored the noise correlation, and derived these values as
\begin{align}
 r_{2t+1} &= \tc \tq_{2t+1} + c (\tc\tU_{2t+1})^2 q_{2t}  
 \label{eq:yanai_r1}\\
 \tr_{2t} &= c q_{2t} + \tc (cU_{2t}^2)^2 \tq_{2t-1}.
 \label{eq:yanai_r2}
\end{align}
In their result, the critical capacity $\alpha_c$ 
is equal to $0.27$, which is not equal to 
our SCSNA analysis and the replica analysis 
($\alpha_c = 0.1998$ for both analyses).
This overestimation comes from the lack of noise correlation evaluation.

In our analysis, 
we consider the effect of crosstalk noise correlation across  $n$-step
previous state, and obtained 
the eqs.(\ref{eq:r_expand1}) and (\ref{eq:r_expand2})
which includes Yanais' analysis (eqs.(\ref{eq:yanai_r1}) and
(\ref{eq:yanai_r2})). 
In the next section, we show that 
the analysis accuracy improves as $n$ in increased ($n = 2, 3, \cdots$).
Hereafter, we call the statistical neurodynamics considering across the
$n$-step previous state as the  ``n-step'' analysis in the following.
``Full-step'' analysis means using all the macroscopic parameters from
the initial state ($t = 0$) to the current state.

%
%


\section{Result}
\label{sec:compare}
\begin{figure*}[t]
 \begin{center}
  \begin{tabular}{ccc}
  \resizebox{0.32\textwidth}{!}{\includegraphics{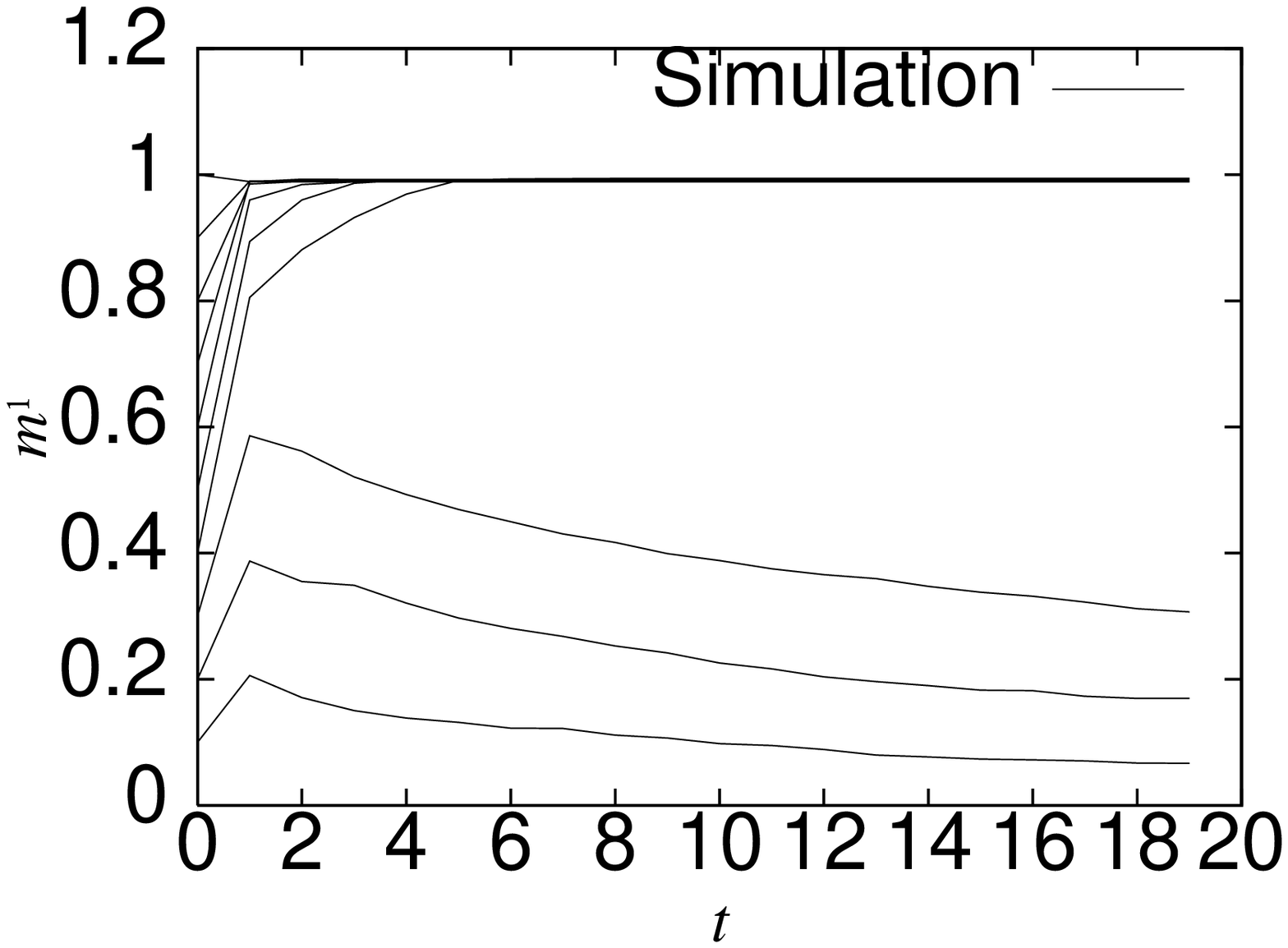}}
   &
  \resizebox{0.32\textwidth}{!}{\includegraphics{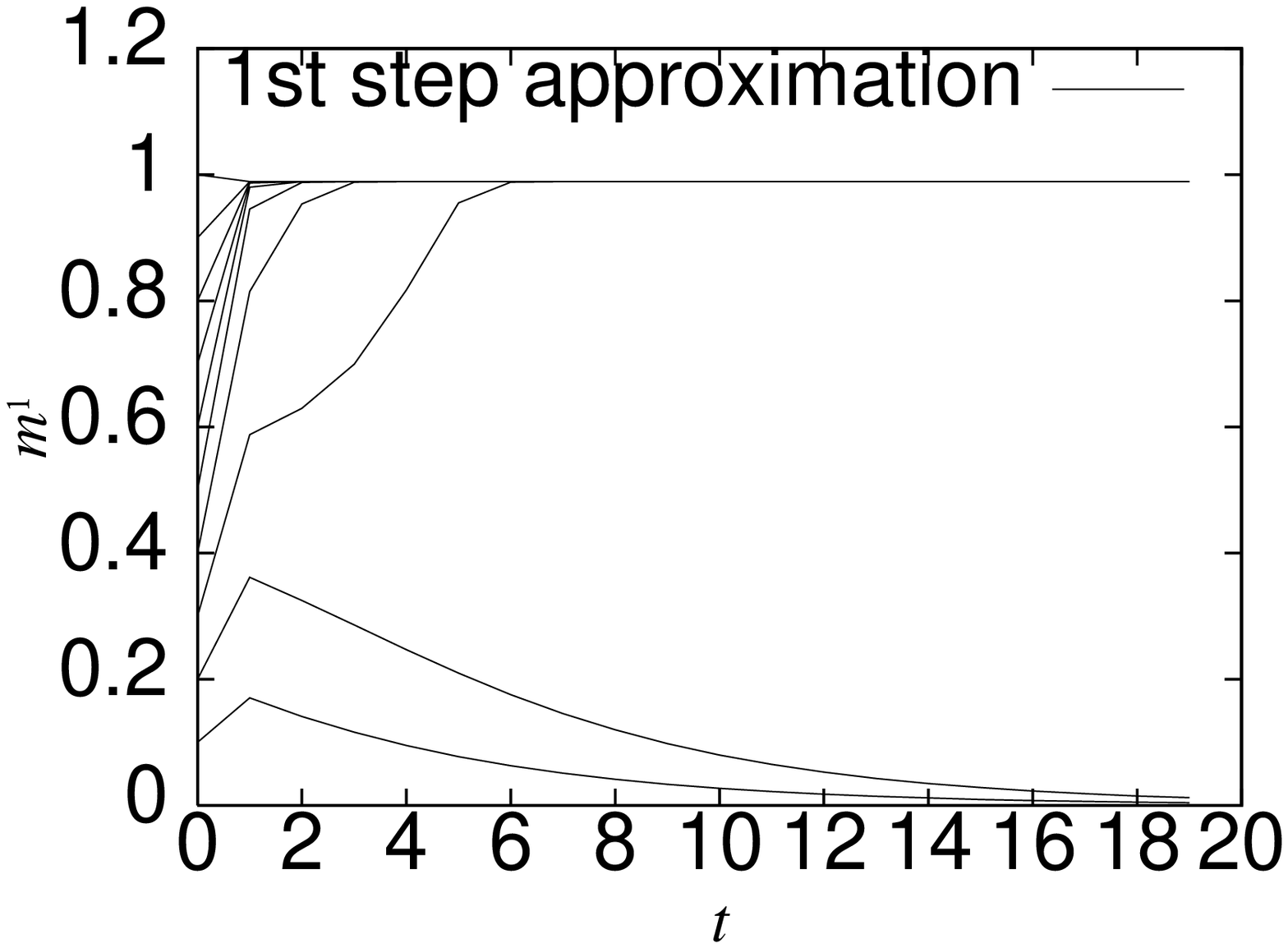}}
   &
  \resizebox{0.32\textwidth}{!}{\includegraphics{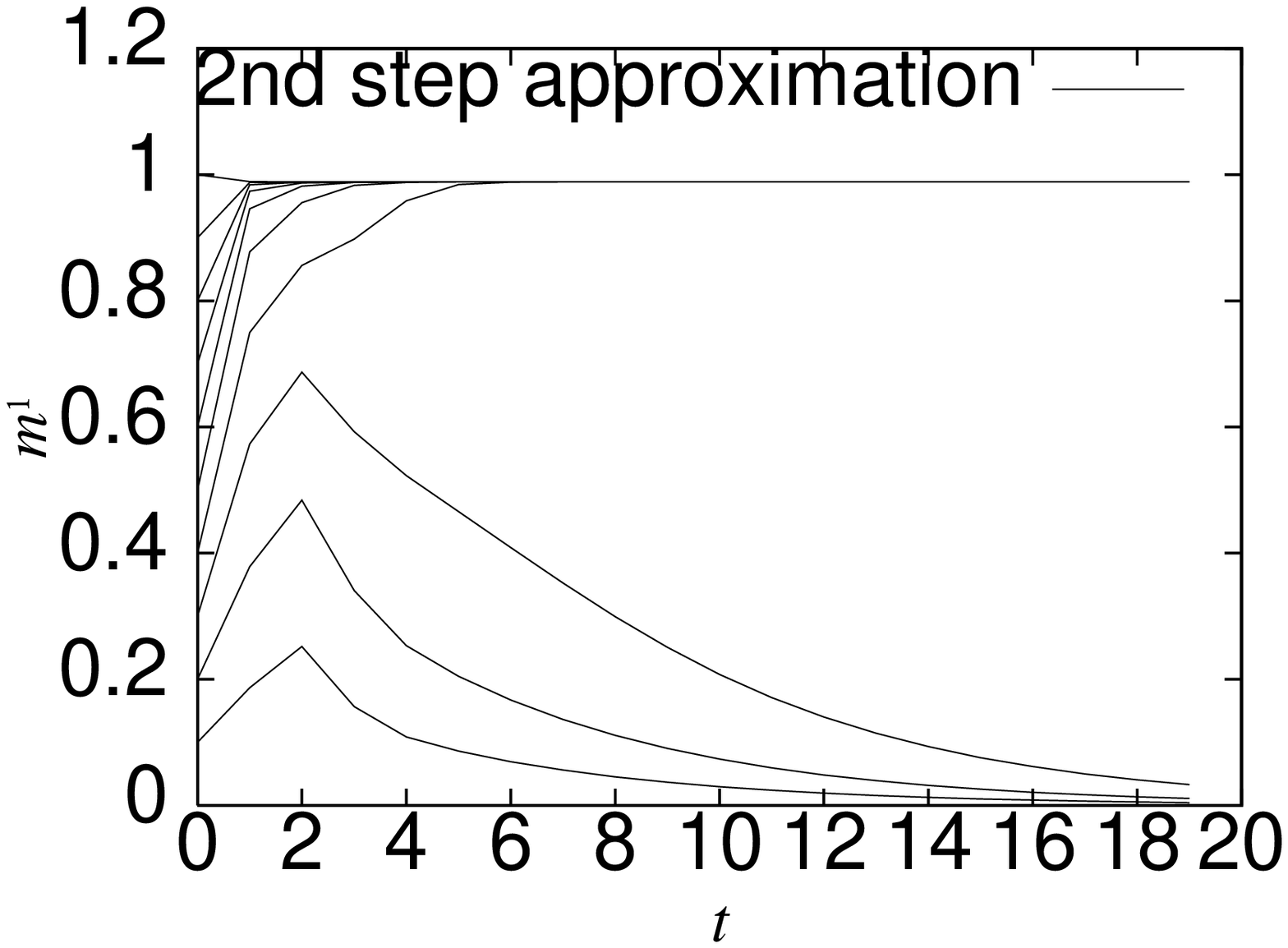}}
   \\
   (a) Simulation
   &
   (b) 1-step analysis
   &
   (c) 2-step analysis
   \\
   &
   &
   \\
   &
   \resizebox{0.32\textwidth}{!}{\includegraphics{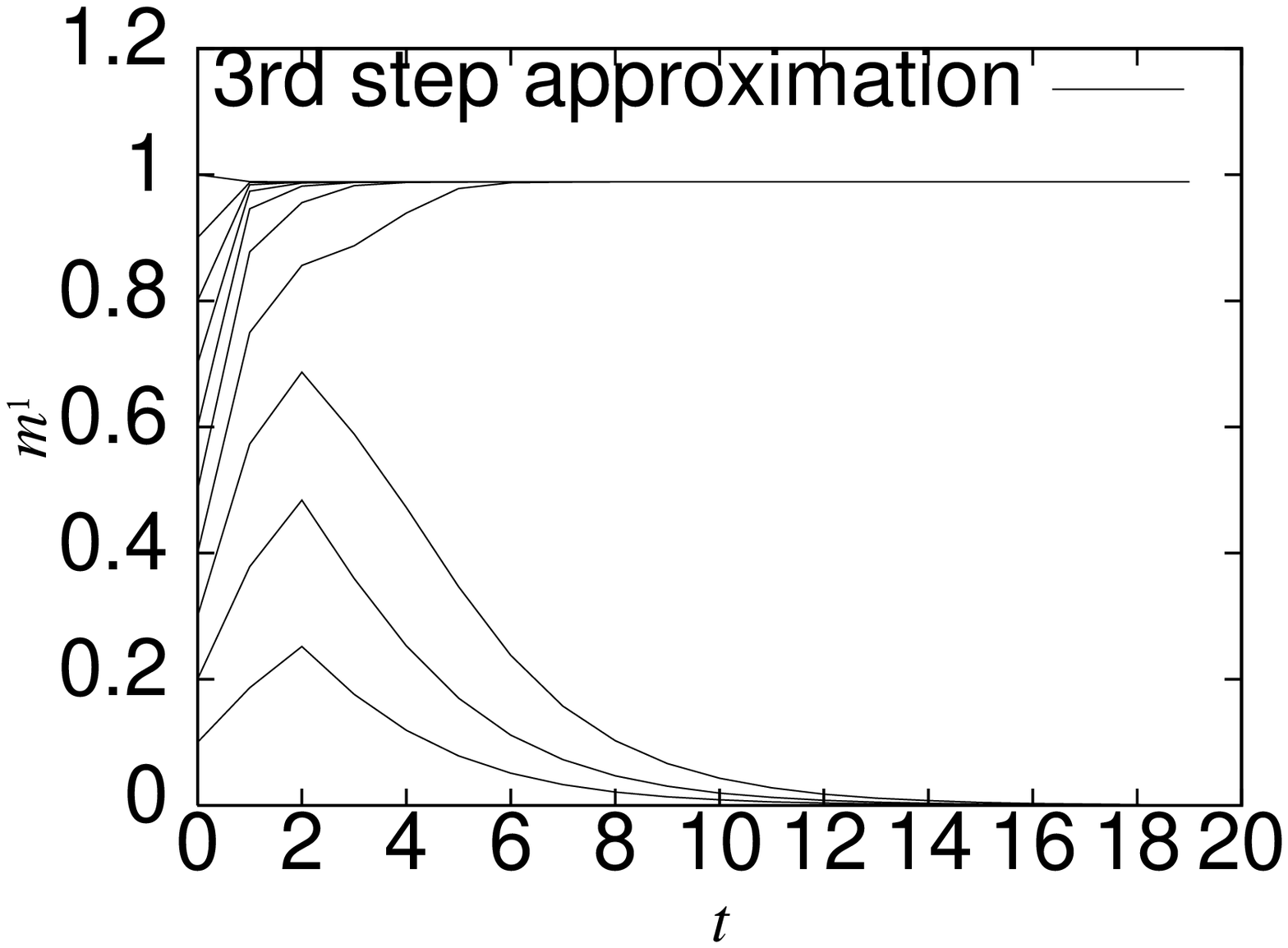}}
   &
   \resizebox{0.32\textwidth}{!}{\includegraphics{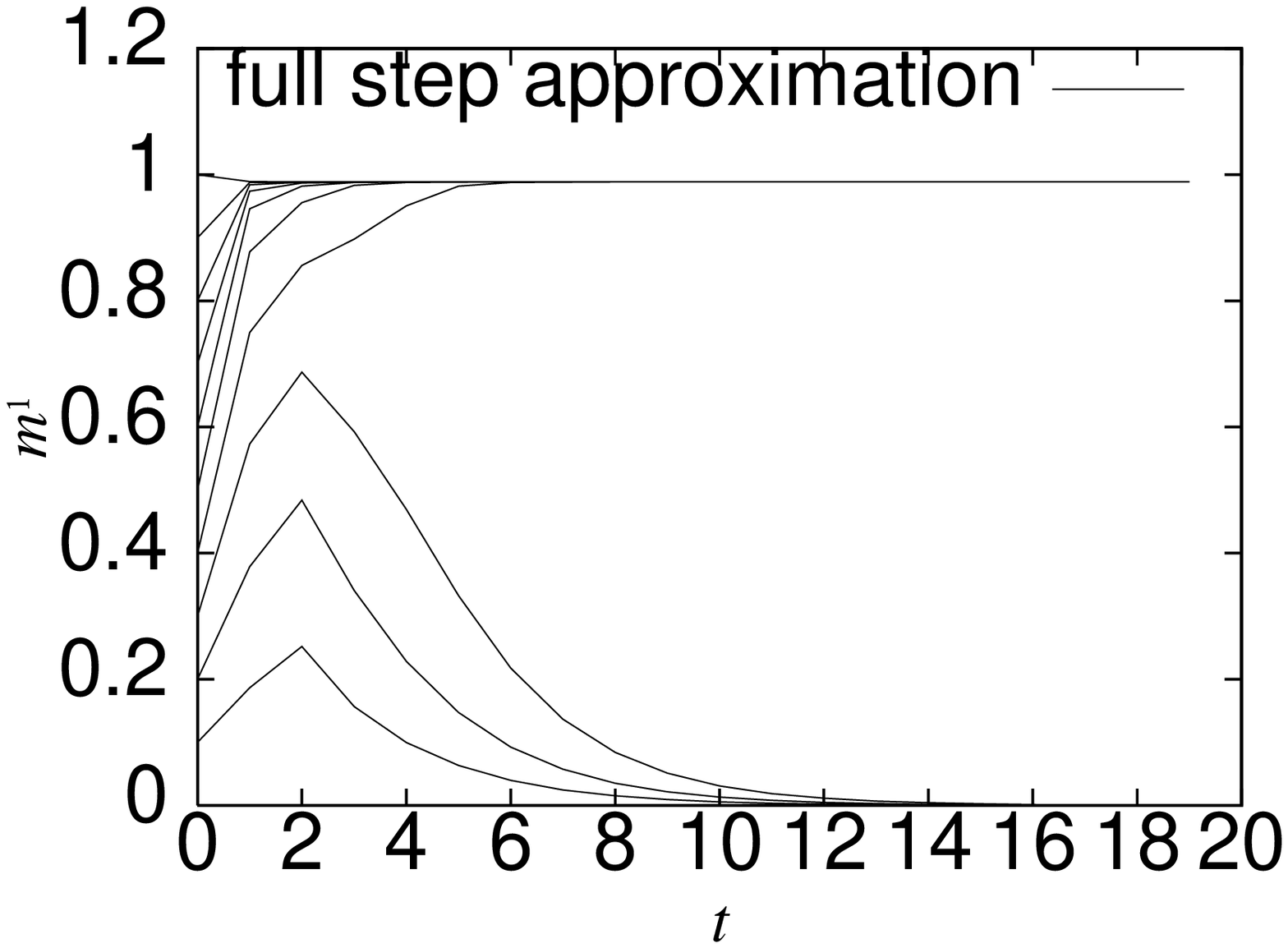}}
   \\
   &
   (d) 3-step analysis
   &
   (e) full-step analysis
  \end{tabular}
 \end{center}
 \caption{Retrival process of a computer simulation and the statistical
 neurodynamics. 
 The horizontal axis means time index $t$, and the vertical axis means 
 the overlap $m$.
 (a) shows a result of computer simulation. From (b) to (e) shows the
 results of statistical neurodynamics. 
 }
 \label{fig:overlap}
\end{figure*}

In this section, 
we compare the  results of the statistical neurodynamics
with those of computer simulation.
%
Fig. \ref{fig:overlap} shows the time evolution  of the overlap $m^{2t}_{1}$,
which means how well the pattern $\Vec{\xi}^{1}$ is retrieved in the 
first layer at $2t$.
Each abscissa axis represents the time step $t$ and 
each ordinate axis  describes overlap $m_1^{2t}$.
Convergence of the overlap $m_1^{2t}$ to $1.0$ means success in retrieving
$\Vec{\xi}^{1}$. 
In the graphs, 
we show several evolution curves in which the initial overlap  
($m_1^{0}$) starts with a different state.

Fig. \ref{fig:overlap}(a)  shows the simulation results. 
We set the number of neurons $N$ to $10,000$, $c = \tc = 1$,
and  the number of stored pattern pairs was indexed as $\alpha=0.15$.
%
The retrieval was successful when we set the initial overlap larger than $0.4$,
and it failed when we set it to $0.3$ or less.
Fig. \ref{fig:overlap}(b) shows the results of the one-step analysis, 
and figs.\ref{fig:overlap}(c)  to (e) shows the result of 
the $2$-step, $3$-step, and full step analysis, respectively.
In each analysis result, the overlap converged to $0$ when retrieving failed
because of assuming infinite neuron units ($N\rightarrow\infty$) exist.
In the simulation results, fig.\ref{fig:overlap}(a), 
the system settled into a spurious memory state when retrieving failed
because the number of units was finite ($N=10,000$).
Therefore, the curves starting at $0.1$ to $0.3$ can be regarded as 
retrieval failures.

As shown in fig. \ref{fig:overlap}(b),  
the one-step analysis says that retrieving is successful when the initial
overlap is $0.3$, which does not agree with the simulation results.
Fig. \ref{fig:overlap}(c) shows the results for $n=2$, {\it i.e.} the 2-step analysis.
The 2-step analysis says that the retrieval is a failure when the
initial overlap is $0.3$, 
which agrees with the simulation results.
Fig. \ref{fig:overlap}(d) and (e) show the 3-step  results
and the full-step analysis results, respectively.
Each figure shows similar characteristics, 
and the results agree with those of the simulation results shown as fig. \ref{fig:overlap}(a).
Since the 3-step analysis results are very similar to the full-step
analysis results 
the 3-step analysis is enough for approximating the full-step analysis.
In other words, the previous 3-step correlations are effective for BAM
retrieval. 

\begin{figure}
 \begin{center}
  \resizebox{0.8\textwidth}{!}{\includegraphics{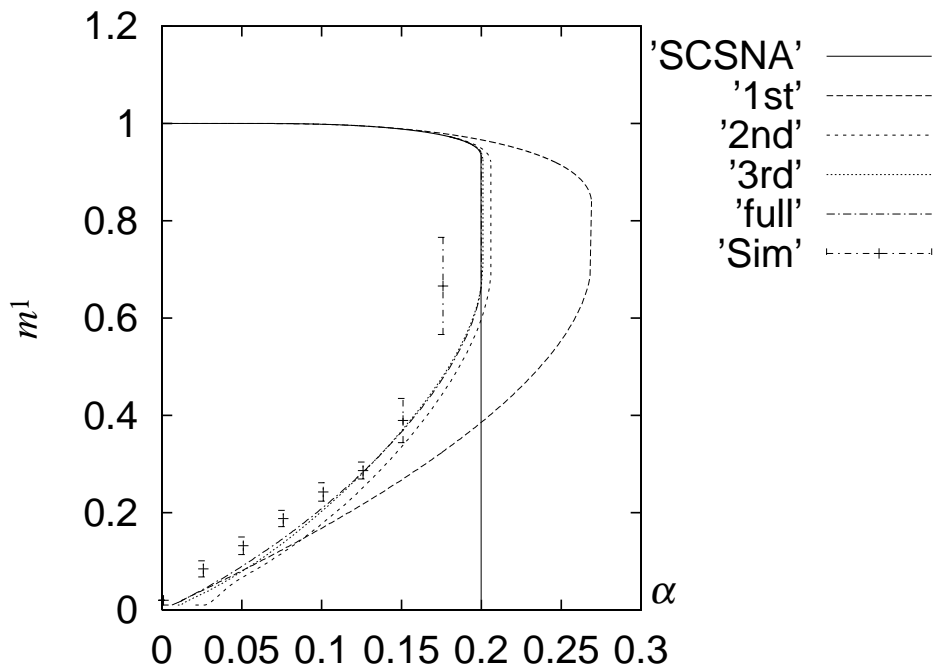}}
 \end{center}
 \caption{Capacity comparing the statistical neurodynamics with
 SCSNA. The horizontal axis means the loading rate $\alpha$, and the
 vertical axis means the overlap $m$.
 The dashed curves shows the results of the statistical neurodynamics.
 The results of computer simulations are shown with error-bar which
 indicates mean with standard deviations.
 }
 \label{fig:basin}
\end{figure}

In the statistical neurodynamics analysis, 
the equilibrium state is described as the limit of the transient process,
and the order parameters should be consistent to the result of the SCSNA.
Fig. \ref{fig:basin} shows the memory capacity and the basin of attraction
%
, which means the degrading limit of the retrievable pattern in the
initial state measured by the overlap $m_1^0$.
%
%
%
%
For example, in fig. \ref{fig:overlap}(d), 
the retrieval is successful when starting at $m_1^{0} = 0.4$, 
while it is a failure when starting at $m_1^{0} = 0.3$.
There is thus a basin when $m_1^{0}$ is between $0.3$ and $0.4$ for 
$\alpha = 0.15$.
%
%
%
The dashed curves in fig.\ref{fig:basin} are derived from the
statistical neurodynamicses. 
In these curves,
the upper part shows equilibrium overlap $m_{1}^{\infty}$ in successful
retrieval and the lower part shows the basin of attraction $m_{1}^{0}$.
When we set the initial overlap $m_{1}^{0}$ to be in the area surrounded
by these curves, the retrieval will be success.
Therefore, the area surrounded by these curves represents 
the successful retrieval area. 
It is clear that the one-step analysis overestimates 
both the relative capacity and the basin of attraction\cite{Yanai91}. 
The theoretical estimation accuracy improves and comes close to that of the
SCSNA analysis asymptotically as the analysis accuracy is improved 
(2-step, 3-step, $\cdots$.)
We also show the basin derived from the simulation results using
error-bars in fig.\ref{fig:basin}. 
The results of simulation agree with those of the statistical
neurodynamics quantitatively.

\section{Conclusion}
\label{sec:conclusion}
We derived the macroscopic parameters of BAM in the equilibrium state by
using the SCSNA and obtained the critical capacity $\alpha_c$ as $0.1998$.
The results agreed with the previous results and the simulation results.

We also analyzed the transient process of BAM using the statistical
neurodynamics and 
obtained the evolution equations for the macroscopic parameters.
Comparison of  the numerical solutions with the simulation results, 
we showed that the analysis results can explain the simulation results
with sufficient accuracy for the transient process.
Therefore,
to explain the transient process of BAM quantitatively,
it is sufficient to consider the 3-step statistical neurodynamics, 
which means that the crosstalk noise has effective correlation across
the 3-step previous state.

\small
\bibliographystyle{unsrt}
\bibliography{shouno}	

\appendix
\section{Detail SCSNA Description}
In Sec.\ref{sec:SCSNA}, we introduced the overlaps, 
\begin{align}
 m_{\mu} &= \frac{1}{cN} \sum_{i=1}^{cN} \xi^{\mu}_i  x_i \\
 \tm_{\mu} &= \frac{1}{\tc N} \sum_{j=1}^{\tc N} \txi^{\mu}_j  \tx_j
\end{align}
for each layer state.
In the equilibrium state, we assumed that the first pattern pair,
($\Vec{\xi}^1$, $\Vec{\txi}^1$), is retrieved, 
so that overlaps for other pattern pairs are small, {\it i.e.}
$m_{\mu}, \tm_{\mu} \sim O(\frac{1}{\sqrt{N}})$ where $\mu \geq 2$.
Thus we denotes the $m_{\mu}$ 
\begin{align}
 m_{\mu} &= \frac{1}{cN} \sum_{i=1}^{cN} \xi^{\mu}_i 
 F( \tc \sum_{\mu=1}^{\alpha N} \xi_i^{\mu} \tm_{\mu}) \\
 &\sim
 \frac{1}{cN} \sum_{i=1}^{cN} \xi^{\mu}_i
 F( \tc \sum_{\nu \neq \mu}^{\alpha N} \xi_i^{\nu} \tm_{\nu})
 + 
 \frac{1}{cN} \sum_{i=1}^{cN} \tc\tm_{\mu}
 F'(  \tc \sum_{\nu \neq \mu}^{\alpha N} \xi_i^{\nu}\tm_{\nu} )
 \\
 &=
 M_{\mu} + \tc \tm_{\mu} U,
 \label{eq:M1}
\end{align}
where
\begin{align}
 U &= \frac{1}{cN} \sum_{i=1}^{cN} F'(\tc \sum_{\nu \neq \mu}^{\alpha N} \xi_i^{\nu}\tm_{\nu})\\
 M_{\mu} &= 
 \frac{1}{cN} \sum_{i=1}^{cN} \xi^{\mu}_i
 F( \tc \sum_{\nu \neq \mu}^{\alpha N} \xi_i^{\nu} \tm_{\nu})
 =
  \frac{1}{cN} \sum_{i=1}^{cN} \xi^{\mu}_i x_i^{(\mu)}.
\end{align}
We denote the $x_i^{(\mu)}$ as the value drawn
the effect of $\mu$th pattern pair from $x_i$, {\it i.e.} $x_i - x_i^{(\mu)} \sim O(\frac{1}{\sqrt{N}})$.
$\tm_{\mu}$ is also denoted
\begin{align}
 \tm_{\mu} \sim \tM_{\mu} + c m_{\mu} \tU,
 \label{eq:M2}
\end{align}
where
\begin{align}
 \tU &= \frac{1}{\tc N} \sum_{i=1}^{\tc N} F'(c \sum_{\nu \neq \mu}^{\alpha N} \txi_j^{\nu} m_{\nu})\\
 \tM_{\mu} &= 
  \frac{1}{\tc N} \sum_{j=1}^{cN} \txi^{\mu}_j \tx_j^{(\mu)}.
\end{align}
Solving eqs.(\ref{eq:M1}) and (\ref{eq:M2}) for $m^{\mu}$ and $\tm^{\mu}$, 
we obtain
\begin{align}
 m_{\mu} &= \frac{1}{1-c\tc U\tU} (M_{\mu} + \tc U \tM_{\mu}), 
 \label{eq:app_ovlp1}\\
 \tm_{\mu} &= \frac{1}{1-c\tc U\tU} (\tM_{\mu} + c \tU M_{\mu}).
 \label{eq:app_ovlp2}
\end{align}
Since the noise terms in eqs.(\ref{eq:eq_update1}) and (\ref{eq:eq_update2}) 
can be described as
$
 z_i = \tc \displaystyle\sum_{\nu \geq 2}^{\alpha N} \xi_i^{\nu} \tm_{\nu}$, 
 and
$ \tz_j = c \displaystyle \sum_{\nu \geq 2}^{\alpha N} \txi_i^{\nu} m_{\nu}$,
we substituted eqs.(\ref{eq:app_ovlp1}) and (\ref{eq:app_ovlp2}) to these noises
and obtained
\begin{align}
 z_i &= \frac{\alpha \tc \tU}{1 - c\tc U\tU} x_i^{(\mu)} + Z_i, \\
 \tz_j &= \frac{\alpha c U}{1 - c\tc U\tU} \tx_j^{(\mu)} + \tZ_j,
\end{align}
where
\begin{align}
 Z_i &= \frac{1}{N(1- c\tc U\tU)}
 \sum_{\nu \geq 2}^{\alpha N} 
 \left(
 \tc \tU \sum_{k\neq i} \xi_i^{\nu} \xi_k^{\nu} x_k^{(\mu)} +
 \sum_{j=1}^{\tc N} \xi_i^{\nu} \txi_j^{\nu} \tx_j^{(\mu)}
 \right)
 \\
 \tZ_j &= \frac{1}{N(1- c\tc U\tU)}
 \sum_{\nu \geq 2}^{\alpha N} 
 \left(
 c U \sum_{l\neq j} \txi_j^{\nu} \txi_l^{\nu} \tx_l^{(\mu)} +
 \sum_{i=1}^{c N} \txi_j^{\nu} xi_i^{\nu} x_i^{(\mu)}
 \right)
\end{align}

We assumed $Z_i$ and $\tZ_j$ as independent identical Gaussian noise, described as 
$Z_i \sim N(0, \alpha r)$ and $\tZ_j \sim N(0, \alpha \tr)$, and evaluated expectations
$E[(Z_i)^2]$ and $E[(\tZ_j)^2]$.
We then obtained
\begin{align}
 E[(Z_i)^2] &= \alpha r = \frac{\alpha \tc}{(1- c \tc U\tU)^2} (c \tc \tU^2 q + \tq), \\
 E[(\tZ_j)^2] &= \alpha \tr = \frac{\alpha c}{(1- c \tc U\tU)^2} (c \tc  U^2 \tq + q),
\end{align}
where
\begin{align}
 q &= \frac{1}{cN}\sum_{i=1}^{cN} (x_i^{(\mu)})^2 \\
 \tq &= \frac{1}{\tc N}\sum_{j=1}^{\tc N} (\tx_j^{(\mu)})^2
\end{align}
From the self-averaging property, 
we obtained the SCSNA order parameter equation in sec.\ref{sec:SCSNA}.

\section{Correlation with n-step previous state }
To evaluate the effect of the previous crosstalk noises,
we must derive the correlation of a unit between the current state and
the $n$-step before state. 
These are described by $\tq_{2t+1,2(t-n)+1}$ and $q_{2t,2(t-n)}$, respectively:
\small
\begin{align}
 &
 \tq_{2t+1,2(t-n)+1} = 
 \notag\\
 &
 \frac{
 \int D\Vec{z}\exp( -\Vec{z}^{\mathrm T} \tSigma^{-1} \Vec{z}) 
 \langle \tY_{2t+1}(z_1) \tY_{2(t-n)+1}(z_2) \rangle }
 {2\pi\left| \tSigma \right| }
\notag\\
 &
 q_{2t,2(t-n)} = 
 \notag\\
 &
  \frac{
 \int D\Vec{\tz}\exp( -\Vec{\tz}^{\mathrm T} \Sigma^{-1} \Vec{\tz}) 
 \langle Y_{2t}(\tz_1) Y_{2(t-n)}(\tz_2) \rangle 
 }
 {2\pi\left| \Sigma \right| }
\end{align}
\normalsize
The matrices  $\Sigma$, $\tSigma$ and vectors $\Vec{z},
\Vec{\tz}$ are described as follows:
\small
\begin{align}
 \tSigma &= 
 \begin{pmatrix}
         \tr_{2t} & \tr_{2t,2(t-n)} \\
  \tr_{2t,2(t-n)} & \tr_{2(t-n)}    \\
 \end{pmatrix}
 \notag\\
 \Vec{z} &= 
 \begin{pmatrix}
  z_1 \\
  z_2
 \end{pmatrix}
 \notag\\
%
%
 \Sigma &= 
 \begin{pmatrix}
  r_{2t-1}      & r_{2t-1,2(t-n)-1} \\
  r_{2t,2(t-n)} & r_{2(t-n)-1} \\
 \end{pmatrix}
 \notag\\
 \Vec{\tz} &=
 \begin{pmatrix}
  \tz_1 \\
  \tz_2
 \end{pmatrix}
 \end{align}
%
The diagonal components of each matrix correspond to the 
variances of the current state noises.
The non-diagonal components express the noise correlation between
the current state and the $\eta$-step previous state.
For $\eta \geq 2$, the correlations are described as:
\small
\begin{align}
 r_{2t+1,2(t-\eta)+1} = & \tc \tq_{2t+1,2(t-\eta+1)+1}
 \notag\\
 &
 +  c\tc \tU_{2t+1} U_{2t} r_{2t-1,2(t-\eta+1)+1},
 \notag\\
 \tr_{2t,2(t-\eta)} = & cq_{2t,2(t-\eta)} 
 \notag\\
 &
 + c\tc U_{2t} \tU_{2t-1} \tr_{2(t-1),2(t-\eta)},
\end{align}
\normalsize
and for $\eta \geq 3$, they are
\small
\begin{align}
 &
 r_{2t+1,2(t-\eta)+1} = \tc \tq_{2t+1,2(t-\lambda)+1}
 \notag\\ 
 &\quad
 + c\tc^2 \tU_{2t+1} \tU_{2(t-\lambda)+1} 
 q_{2t,2(t-\lambda)} 
 \notag \\ 
 &\quad
 + \tc \sum_{\lambda=1}^{n} 
 (c\tc)^\lambda \tq_{2(t-\lambda)+1,2(t-\eta)+1} 
 \!\!\!\! \prod_{\tau=t-\lambda+1}^{t} 
 \!\!\!\! \tU_{2\tau+1} U_{2\tau}
 \notag \\
 &\quad
 + \tc
 \sum_{\lambda=1}^{n-\eta} 
 (c\tc)^\lambda \tq_{2t+1,2(t-\eta-\lambda)+1} \!\!\!\!\!\!
 \prod_{\tau=t-\lambda+1}^{t} \!\!\!\!\!\!
 \tU_{2(\tau-\eta)+1} U_{2(\tau-\eta)}
 \notag \\
 &\quad
 + c\tc^2 \tU_{2t+1} \tU_{2(t-\eta)+1}
 \notag \\
 &\qquad\qquad
 \sum_{\lambda=1}^{n-1}(c\tc)^{\lambda}
 q_{2(t-\lambda),2(t-\eta)}
 \prod_{\tau=t-\lambda+1}^{t} U_{2\tau} \tU_{2\tau-1}
 \notag \\
 &\quad
 + c\tc^2 \tU_{2t+1} \tU_{2(t-\eta)+1}
 \sum_{\lambda=1}^{n-1-\eta}(c\tc)^{\lambda}
 q_{2t,2(t-\eta-\lambda)}
 \notag \\ 
 &\quad\qquad
 \prod_{\tau=t-\lambda+1}^{t} U_{2(\tau-\eta)} \tU_{2(\tau-\eta)-1}
 \notag \\
 &\quad
 + (c\tc)^2 \tU_{2t+1} U_{2t} \tU_{2(t-\eta)+1} U_{2(t-\eta)}
 r_{2t-1,2(t-\lambda)-1}
 \notag
\end{align}
\begin{align}
 &
 \tr_{2t,2(t-\eta)} =   cq_{2t,2(t-\eta)}
 \notag\\ 
 &\quad
 + \tc c^2 U_{2t} U_{2(t-\eta)} 
 \tq_{2t-1,2(t-\eta)-1} 
 \notag \\
 &\quad
 + 
 c \sum_{\lambda=1}^{n} (c\tc)^\lambda 
 q_{2(t-\lambda),2(t-\eta)} 
 \!\!\!\! \prod_{\tau=t-\lambda+1}^{t} 
 \!\!\!\! U_{2\tau} \tU_{2\tau-1}
 \notag \\
 &\quad
 + 
 c \sum_{\lambda=1}^{n-\eta} (c\tc)^\lambda 
 q_{2t,2(t-\eta-\lambda)} \!\!\!\!
 \prod_{\tau=t-\lambda+1}^{t} \!\!\!\! 
 U_{2(\tau-\eta)} \tU_{2(\tau-\eta)-1}
 \notag \\
 &\quad
 + {\tc}c^2 U_{2t}U_{2(t-\eta)}
 \notag\\ 
 & \qquad\qquad
 \sum_{\lambda=1}^{n-1}(c\tc)^{\lambda}
 \tq_{2(t-\lambda)-1,2(t-\eta)-1}
 \!\!\!\!\!\!
 \prod_{\tau=t-\lambda+1}^{t} 
 \!\!\!\!\!\!
 \tU_{2\tau-1} U_{2\tau-2}
 \notag \\
 &\quad
 +
 {\tc}c^2 U_{2t}U_{2(t-\eta)}
 \sum_{\lambda=1}^{n-1-\eta}(c\tc)^{\lambda}
 \tq_{2t-1,2(t-\eta-\lambda)-1}
 \notag \\
 &\quad\qquad
 \prod_{\tau=t-\lambda+1}^{t} \tU_{2(\tau-\eta)-1} U_{2(\tau-\eta)-2}
 \notag \\ 
 &\quad
 +
 (c\tc)^2 U_{2t} \tU_{2t-1} U_{2(t-\eta)} \tU_{2(t-\eta)-1}
 \tr_{2(t-1),2(t-\lambda-1)}
 \notag\\
\end{align}
\normalsize

\newpage

\end{document}